\documentclass[twocolumn,showpacs,preprintnumbers,amsmath,amssymb,prl]{revtex4}

\usepackage{graphicx}
\usepackage{dcolumn}
\usepackage{bm}
\usepackage{subfigure}

\begin{document}


\title{Geometry-induced frustration of magnetization
in a planar soft-hard magnetic system}

\author{Hemachander Subramanian and J. E. Han}
\affiliation{
Department of Physics, State University of New York at Buffalo, Buffalo, NY 14260, USA}
  
\date{\today}

\begin{abstract}
We computationally study the frustrated magnetic configurations of a
thin soft magnetic layer with the boundary condition fixed by
underlying hard magnets. Driven by geometrical constraints and
external magnetic field, transitions between frustrated energy
minima result in magnetic hysteretic behavior. The presence of
soft-magnet introduces strong undulations in the energy
landscape in a length scale set by the magnetic property of the soft magnet.
We propose a possible use of the phenomena to locally control
the movement of magnetic nanoparticles. 
\end{abstract}


\maketitle

Magnetic systems with geometrical frustration has attracted
considerable attention for the presence of multiple ground
states and intricate thermodynamics properties~\cite{moessner,
schiffer} for many decades.  Soft-hard magnetic structures
have recently been intensively investigated for their high
coercivity and saturation magnetization properties
~\cite{kneller, coehoom, zeng}. Exchange-coupled planar
soft-hard magnetic structures ~\cite{eiichi, sabir1, sabir2,
asti} show a rich variety of magnetic configurations when
magnetic frustration is induced geometrically. This is
particularly the case when the characteristic length of
magnetization configuration is comparable to the device size.

Previous works on soft-hard magnetic systems~\cite{siegmann,
rohl, hellwig, stamm} have mostly focussed on magnetization
configurations between layers of soft-hard magnets.  In this
paper, we study two-dimensional geometry-induced multiple magnetic
configurations, its associated configurational energy minima,
and external magnetic field-induced transitions between these
configurations. We concentrate on the magnetization configurations in
the plane of the soft magnet, determined by the magnetization
directions of the constraining hard magnets and an external
field. 

Controlling magnetic states has a wide range of applications 
such as memory devices, sensors that use magnetostriction, etc. Here we focus on a
magneto-mechanical application, namely, moving 
magnetic nanoparticles. The ability to move,
confine, and rotate nanoparticles has come from creating
non-uniform magnetic fields using either hard magnets or
electromagnets ~\cite{gosse, devries, yan, trepat, mirowski}.
We use geometrical frustration to create localized non-uniform magnetic fields. 
Physical properties of biomolecules such as tensile strength
can be studied by attaching magnetic nanoparticles to them and
applying force on the nanoparticles by trapping them in such localized
magnetic energy minima.

\begin{figure}  
\includegraphics[width=2.1in]{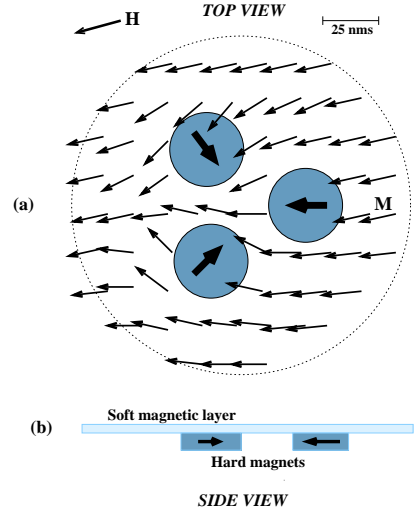}
\caption{(a) 2-dimensional soft-hard magnetic composite 
system. The magnetic structure in the soft-magnetic region is
determined by the boundary condition fixed by the hard
magnetic discs and external field ${\bf H}$. One of possible
energy minima, with two magnetic in-flux and one out-flux
between the discs, is shown. (b) Side-view of the device.}
\end{figure}

The 2-dimensional soft-hard magnetic system under study
consists of three hard magnetic discs placed underneath a thin soft
magnetic layer (see FIG. 1). The three hard magnets, placed on
vertices of an imaginary equilateral triangle, are chosen to
have their magnetizations pointing toward the center of the
triangle. The role of the hard magnets is to fix the
magnetization boundary condition on the soft-magnet at the
interface, and to introduce frustration in the system. We
assume that the magnetization of the soft-magnetic layer is
in-plane, and that the soft magnetic layer is thin enough ($5$
nm) that its magnetization direction varies little across the
width of the layer. FIG. 1 shows one of the possible energy
minimum configurations of the soft-hard magnetic system. 

The energetically favorable configurations are found to be the
ones where the magnetization flows in through two of the necks
(region between two hard magnetic discs) which are then
balanced out by the out-flux through the remaining neck.  Such
a configuration is shown in FIG. 1(a).  Although the
configuration in FIG. 1(a) is at a local energy minimum, the
high energy density near the neck of out-flux due to the high
gradient of magnetization makes the system unstable against
external field and triggers a transition to another local
minimum. Configurations only with in-flux or out-flux are
highly unfavorable energetically due to the high divergence of
magnetization at the center of the soft-hard magnet, but will
nevertheless be reached if the system's initial
magnetization is closer to the above mentioned configurations. 

Even with the magnetization restricted to be in-plane, 
the above-mentioned unfavorable configuration can be made stable.
Introduction of a hole (or non-magnetic region) at the center
of the soft
magnetic disc allows magnetization to form a vortex around the
hole, thereby avoiding high energy cost. The other way is to
let the magnetization to point along z-direction by
controlling the thickness of the magnetic layer, in which
case, the energy cost of convergence will be minimized
considerably, with the magnetization choosing to point in
z-direction at the point of convergence. Such deviations could
be utilized in devices to manipulate the softness of
magnetization.

The system of FIG. 1 can be fabricated using CoCr-based alloy as hard
magnetic discs and permalloy or FeCo as soft magnet.
CoCr-alloy thin film can be grown by sputtering and patterned
lithographically into three circular discs forming an
equilateral triangle. MFM tip can be used to pin the
magnetization orientation of the discs to predefined
directions.  Soft magnetic film can then be coated on top of
the hard discs.

The equation for Hamiltonian density of the system is
\begin{equation}
{\cal H} = \frac{J}{3A^2}|\nabla {\bf m}|^2-M_0^2{\bf m}\cdot{\bf h}
+V_0(|{\bf m}|^2-1)^2 
\end{equation}
in CGS units. 
Here, $A$ is the distance between two lattice points in the discrete mesh
created to simulate the equation,  
${\bf m}$ the magnetization vector normalized to the
saturation magnetization $M_0$, $J$ the
exchange-energy coefficient, ${\bf h}$ the applied magnetic field
normalized to $M_0$, and $V_0$ the confinement potential
to impose the condition of $|{\bf m}|=1$ ~\cite{han}.
We discretize the energy density ${\cal H}$ 
on the soft-magnetic region on a triangular mesh system. 

The first term is the exchange term, approximated for a
continuous dipole distribution~\cite{kittel}. The second term
is just the energy of a dipole in an external magnetic field
${\bf h}$. The last term with a large value of $V_0$
is concocted to preserve $|{\bf
m}|^2=1$ condition during numerical energy minimization. 
Inclusion of the condition $|{\bf m}|^2=1$ in the energy
functional had better numerical behaviors than a direct
imposition of the constraint.
The resulting energy minima satisfied the condition $|{\bf
m}|^2=1$ accurately. 

We assign the values of bulk Nickel for $J$ and $M_0$, $A = 5$
nm, $V_0=50$, $|{\bf H}| = M_0|{\bf h}| = 250$ Oersted
~\cite{han}. We let the external magnetic field rotate around
the system to drive the transitions, and evaluated the minimum
energy configurations for all external field directions.
Typical values of radius of hard magnets, distance between
hard magnetic centers and the radius of the soft magnetic
circle are $20, 60$ and $150$ nm respectively. 

To keep the analysis of the soft-hard magnetic system simple, only
the exchange-interaction in the soft-magnetic region
was considered for the energy minimization calculation of the
soft-magnet. Other magnetic interactions, especially 
the dipole-dipole interaction and the magnetocrystalline 
anisotropy energy, were left out. The effect of anisotropy energy 
can be made small by choosing low crystalline anisotropy 
materials, such as permalloys. One of the main 
results of this work is to show that the geometrical constraint 
leads to many equivalent local energy minima states in the soft-magnetic 
region and we do not expect that this general conclusion will change 
by including other interactions.

The system was simulated on a triangular lattice of size
$50\times 50$. For energy minimization, we used the conjugate gradient
method~\cite{payne}. Polak-Ribiere method was used in updating
the search directions for minimization~\cite{press}.
We have checked that our data does not change
significantly with larger mesh systems.  The soft magnet is
taken to be circular to preserve the three-fold symmetry of
the system, although the symmetry need not be exact. 
Since we keep only the exchange coupling, the  
demagnetization effects from the boundary would be absent. 
We used free boundary conditions. The energy is concentrated
near the 3-disc region, well separated from the boundary points,
justifying our choice of boundary conditions. 

\begin{figure}  
\includegraphics[width=2.4in]{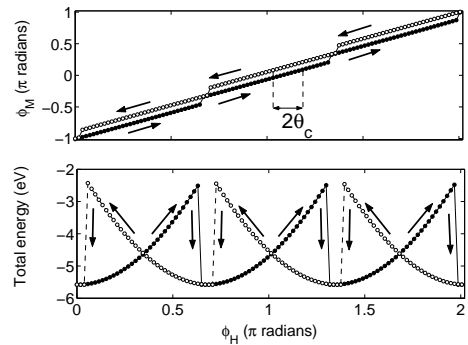}
\caption{(a) The angle of the magnetization averaged over the
soft-magnetic region $\phi_M$  as a function of the external field
direction $\phi_H$. The coercivity angle $\theta_c$ is defined
as half of the difference in $\phi_H$
between the paths of clock-wise and
counter-clock-wise rotations of the external field, at a given $\phi_M$.
(b) Total energy of the system. At the transition the system makes
transition to another local minimum of lower energy. The sense of rotation of the
external field is shown by arrows.}
\end{figure}

The variation of total energy and the angle of average
magnetization of the soft-hard magnet with the external 
field's angle is shown in FIG. 2.  
Three hysteresis transitions happen within one full rotation of external
magnetic field. These hystereses correspond to the switching
of the system between the three equivalent minimum energy
states. The energy drop during the transition is a function of 
the geometrical parameters of the system and the 
magnitude of applied field. 

The regular definition of coercive field does not apply in our
case because the angle, and not the magnitude of the applied
field, is varied. The same magnetic configuration can be
reached either by a clockwise or by a counter-clockwise rotation of the
external magnetic field.  Because of the presence of
hysteresis in the system, these two angles would not be the
same.  We define our coercive field angle, $\theta_c$, to be half of the
difference between these two external field angles [FIG. 2(a)].  For the
geometrical parameters of the system mentioned above, the
coercive field angle is about $0.15$ radians for an external
field of $250$ Oersteds.  As the external field intensity is
increased, the energy drop during transition decreases,
because of the earlier onset of collapse of current
configuration to another energetically favorable one.  For our
sample geometry, $H_c$, the critical field below which there
are no transitions, is approximately $100$ Oersteds. 

The scale of variation of magnetization on the plane of the
soft magnet is characterized by only one parameter in this
system: $J$, the exchange coefficient.  An increase in $J$ is
equivalent to increasing the length scale of magnetization and
reducing the scale of sample geometry. Therefore, increasing
$J$ increases the energy drop 
during transition from one minimum to another. 
This is observed in our system. Similarly, $H_c$ scales with $J$. 

Now we discuss the application of magnetic hysteresis to
moving a free magnetic nanoparticle in the device sketched in
FIG. 1(a).  Because of the complex magnetic dipole distribution
of the soft-hard magnetic system that arises due to the
interaction between hard and soft magnets, the energy
experienced by a point magnetic dipole just above the
soft-hard magnet due to the magnetic field created by the
soft-hard magnet exhibits multiple minima
on the surface. The minima are located at high magnetic field
intensity positions, since the energy experienced by the
mobile nanoparticle equals $-\bm{\mu} \cdot [{\bf B}(x,y,z)  +
{\bf H}]$, where ${\bf B}(x,y,z)$ is the magnetic field at a
point $(x,y,z)$ above the soft-hard magnet created by the
magnetic dipoles present in the soft-hard magnet , ${\bf H}$
is the externally applied magnetic field, and ${\bf \mu}$ is
the total magnetic moment of the magnetic nanoparticle,
measured in the CGS units of magnetic moments/cm$^3$.  We
assume that the magnetic nanoparticle's magnetic moment aligns
with the direction of the magnetic field at its position.  The
nanoparticle chooses points where $|{\bf B}(x,y,z) + {\bf H}|$
is maximum.  FIG. 3 shows snapshots of movement of magnetic
energy minima due to the rotating external magnetic field, $z
= 10$ nm above the soft-hard magnet.

\begin{figure}  
\includegraphics[width=2.7in]{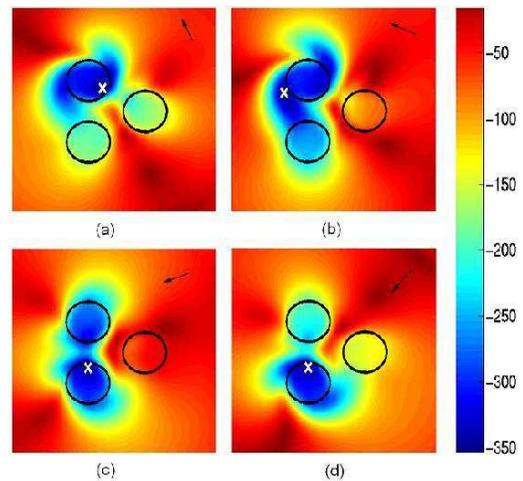}
\caption{(a-d) This sequence of figures show the movement of
magnetic energy minima as experienced by a small nanomagnet on
top of the soft-hard magnet, as the external magnetic field of 
strength $250$ Oersteds is
rotated around, counter-clockwise. Dark arrows on top right of
the figures denote the direction of the externally applied
magnetic field. White $\times$ mark denotes the position of
the nanomagnet on the surface. The movement of the prominent
minimum from the top-left to the bottom-left hard magnetic
circle only is shown.  The colorbar shows the energy
experienced by a $125$ nm$^3$ sized Nickel magnetic
nanoparticle of magnetization $M_0 = 0.53 \times 10^3$
magnetic moments/cm$^3$, in units of $10^{-1}$ meV.}
\end{figure}

If a nanomagnet, of dimensions much smaller than that of the
system under consideration, is allowed to move freely on the
surface of the soft-hard system,  it will find the nearest
magnetic energy minimum and will move along with it when the
magnetic dipole configuration of the underlying soft-hard
magnet changes. The movement of a nanomagnet in one such
prominent energy minimum, due to the rotating magnetic field,
is shown in FIG. 3.

\begin{figure}  
\includegraphics[width=2.2in]{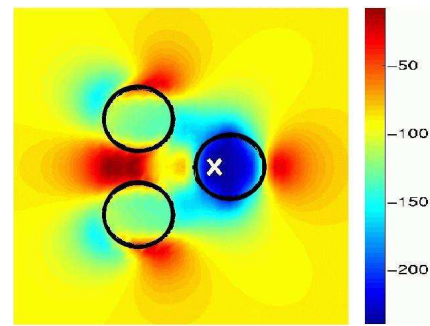}
\caption{Magnetic energy minima as experienced by the nanomagnet
on top of hard magnetic discs without
the soft magnetic medium. All other conditions remain the same as that
of FIG. 3.}
\end{figure}

Soft-hard magnetic surfaces are more efficient in moving
nanoparticles than just hard magnetic surfaces (three discs
without soft magnetic surface).  This is because of the fact
that magnetic energy minima can occur over soft-magnetic
surface too in the former.  The difference between the two can be quantified
by comparing the profiles of energy minima in both these
surfaces. First, the energy minima are deeper in soft-hard
magnet than that of just hard magnetic surface. Typical values
of energy minima would be $-35$ meV for the former, and $-25$
meV for the latter case.  The higher magnetic field strength,
and hence the high magnitude of energy at the minimum, is
because of added soft-magnetic moments.  Second, the curvature
of the minima is more pronounced in soft-hard magnets.
Typical values are around $0.2$ meV/nm$^2$ for the soft-hard
magnetic surface, and $0.08$ meV/nm$^2$ for hard magnetic
surface.  This pronounced curvature traps the nanoparticle
better, and gives us more control on the position of the
nanoparticle. Finally, the size of the potential barrier that
separates two adjacent minima is also more pronounced in
soft-hard magnetic surface compared to hard magnetic surface.
FIG. 4 shows the magnetic field minima on top of hard magnetic
discs without the soft magnetic medium and illustrates the
above mentioned differences. This system does not show any
hysteretic behavior with trapped local minima. It has to be noted
that the range of energies, as shown by the colorbar in FIG. 4
is narrower than the corresponding range in FIG. 3. 

Because of the small size of the nanoparticle, thermal
fluctuations play a role in determining the position of the
nanoparticle, particularly when the multiple minima are
separated by barriers of heights of the order of thermal
energy. Our model includes the effects of thermal fluctuations
on the nanoparticle.  Area enclosed by a contour of energy
$k_BT$ over and above the energy at the current position of
the nanoparticle is accessible to the nanoparticle. The
particle's position would be the deepest minimum within the
area of the contour.  The schematic path of the nanoparticle
during the full rotation of external magnetic field by $2
{\pi}$ radians, is shown in FIG. 5. The orbit of the
nanoparticle is closed. 


\begin{figure}  
\includegraphics[width=1.8in]{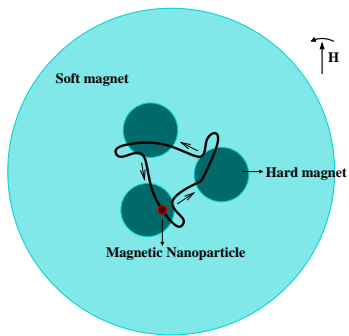}  
\caption{Schematic diagram of the path of a magnetic
nanoparticle on top of soft-hard magnet.  The nanoparticle
traces a path similar to this when externally applied magnetic
field, in the plane of soft-hard magnet, is rotated around the
soft-hard magnet.  The arrows show the direction of movement
of the nanoparticle, when external magnetic field is rotated
counter-clockwise.} \end{figure}

We have investigated a soft-hard magnetic system with a novel
planar geometry and its transitions between frustrated
local energy minima. Interplay between geometrical
constraints imposed by hard magnets and the exchange coupling
inside soft magnet results in the display of complex magnetic
patterns. We present that this rich magnetic structure can be
utilized in local control of magnetization which in turn
enables manipulation of position of magnetic particles. A promising direction
for future work is to use an array of hard magnetic discs with their magnetizations
pinned to introduce frustration, thereby creating magnetic configurations
with periodicity larger than that of the unit cell of the array.

We thank Hao Zeng for helpful discussions. 
This work has been supported by National Science Foundation DMR-0426826.

\end{document}